\documentclass[10pt,conference]{IEEEtran}
\IEEEoverridecommandlockouts
\usepackage[utf8]{inputenc}
\usepackage[T1]{fontenc}
\usepackage[english]{babel}
\usepackage[style=ieee,sortcites=true,sorting=none,maxbibnames=6, backend=biber,mincitenames=1,maxcitenames=2, dashed=false]{biblatex}
\usepackage{graphicx}
\usepackage{amsmath}
\usepackage{amssymb}
\usepackage{amsthm}
\usepackage[autostyle]{csquotes}
\usepackage{mathtools}
\usepackage{braket}
\usepackage{hyperref}
\usepackage{titlesec}
\usepackage{balance}
\usepackage{booktabs}
\usepackage{tabularx}
\usepackage{float}
\usepackage{tikz}
\usepackage{fontawesome}
\usepackage[skip=2pt]{caption}
\usepackage{subcaption}
\usepackage{fancyhdr}
\usepackage{float}
\usepackage{placeins}
\usepackage{cleveref}
\usepackage{algorithm}
\usepackage{algpseudocode}
\usepackage{multirow}
\usepackage{textcomp}
\usepackage[separate-uncertainty=true]{siunitx}

\usepackage{array}

\newcolumntype{L}[1]{>{\raggedright\arraybackslash}p{#1}}

\newcolumntype{R}[1]{>{\raggedleft\arraybackslash}p{#1}}

\newcolumntype{C}[1]{>{\centering\arraybackslash}p{#1}}

\graphicspath{{plots}}
\addto\extrasenglish{

}
\renewcommand{\vec}[1]{\boldsymbol{\mathbf{#1}}} 
\DeclareMathOperator*{\argmin}{argmin}

\titlespacing\section{0pt}{6pt plus 4pt minus 2pt}{4pt plus 2pt minus 2pt}
\titlespacing\subsection{0pt}{6pt plus 4pt minus 2pt}{4pt plus 2pt minus 2pt}
\titlespacing\subsubsection{0pt}{6pt plus 4pt minus 2pt}{4pt plus 2pt minus 2pt}

\newcommand{\code}[1]{$\textbf{\texttt{#1}}$}

\title{The QuaST Decision Tree: Achieving Automation With Data-Based Recommendations \thanks{This project was supported by the Federal Ministry for Economic Affairs and Climate Action on the basis of a decision by the German Bundestag through the project \emph{Quantum-enabling Services and Tools for Industrial Applications (QuaST)}. The project is also supported by the Bavarian Ministry of Economic Affairs, Regional Development and Energy with funds from the Hightech Agenda Bayern.}}

\author{
    \IEEEauthorblockN{Benedikt Poggel\IEEEauthorrefmark{1}, Lena Tokuhiro\IEEEauthorrefmark{1}, Georg Kruse\IEEEauthorrefmark{2}, Jeanette Miriam Lorenz\IEEEauthorrefmark{1}}
    \IEEEauthorblockA{\IEEEauthorrefmark{1}Fraunhofer Institute for Cognitive Systems IKS \\ \IEEEauthorrefmark{2} Fraunhofer Institute for Integrated Systems and Device Technology IISB}
     \{benedikt.poggel, lena.tokuhiro, jeanette.miriam.lorenz\}@iks.fraunhofer.de, georg.kruse@iisb.fraunhofer.de
}
\date{2026-03-30}
\usepackage{fancyhdr}

\fancypagestyle{specialfooter}{%
  \fancyhf{}
  
  \fancyfoot[R]{ \noindent\fbox{%
    \parbox{\textwidth}{%
        {\footnotesize This work has been submitted to the IEEE for possible publication. Copyright may be transferred without notice, after which this version may no longer be
accessible.
}
        }
    }}
}

\begin{document}
\maketitle

\begin{abstract}
Quantum computers are increasingly powerful. Software tools for the development of quantum-enhanced algorithms are maturing. However, the software stack still lacks the connection to applications that would enable hybrid algorithms combining classical and quantum computing steps. End users need to be assisted in choosing the best combination of preprocessing, postprocessing, classical and quantum algorithms options. The application-facing software stack is therefore required to cover problem modeling, encoding, algorithm selection and hyperparameter tuning. A variety of tools exist for specific recommendations. The QuaST Decision Tree reflects the complexity in combining individual decisions in its modular network structure, consisting of flexible computation nodes with modular recommendations. It can easily be  configured to serve in an industrial solver, an HPC software stack, or for rapid prototyping in development. The key ingredient, automation, is delivered by modules. We present one such module judging the feasibility of variational algorithms based on a robust scalability analysis and classification of problem instances. The automation improves the performance of end-to-end solutions, highlights the benefit to be gained from the hybrid quantum solution, reduces expensive trial-and-error testing, and leads to an improved utilization of quantum devices for a practical benefit. 
\end{abstract}

\begin{IEEEkeywords}
quantum computing, hybrid quantum-classical algorithms, applied optimization, abstraction layer, automation, application-facing software stack
\end{IEEEkeywords}

\thispagestyle{empty} 
\pagestyle{plain}
\thispagestyle{specialfooter} 

\section{Introduction}

With quantum computing (QC) maturing, an increasingly relevant practical question is how to leverage the technology for the benefit of industrial applications. In particular, combinatorial optimization (CO) is enormously relevant to many industries such as logistics, production, and finance. Classical computing solutions typically employ a mixture of data preprocessing, exact algorithms and heuristics to tackle the high problem complexities~\cite{weinand_research_2022}, with the recent addition of reinforcement learning methods~\cite{mazyavkina_reinforcement_2021}. Hope for further improvement lies in QC with its inherent ability to represent large search spaces and to use quantum effects for a parallelized search, with advantages in solution quality (e.~g., approximation ratios), time-to-solution and resource efficiency~\cite{abbas_challenges_2024}. 

However, using QC for optimization is no small challenge: Many algorithms promise only a polynomial advantage in computational complexity, bounded by Grover's quadratic speedup~\cite{bennett_strengths_1997}. Thus, for a real-world advantage, careful algorithmic design is needed, accounting for overheads from error correction, latencies, and iterations between classical and quantum computing. Even from the start, it must be determined which \emph{problem instances} classical algorithms struggle with since a general advantage for entire classes of NP-hard problems is out of reach~\cite{abbas_challenges_2024}. 

Moreover, applying quantum algorithms to optimization problems requires \emph{hybrid algorithms} with complementing classical and quantum processing steps. The classical computer is not only required for data preprocessing and solution extraction in postprocessing, but also for decomposition steps needed to handle industrially relevant problem sizes with $10^4$ and more variables. They divide the problem into parts to be treated classically and parts that benefit from QC. In this setting, QC solves specialized tasks just as a graphics processing unit (GPU) would do in a traditional setup. Finally, typical quantum algorithms themselves often require a quantum-classical iteration rather than the execution of a mere quantum circuit, e.g., variational quantum algorithms~\cite{cerezo_variational_2021}, or Quantum Branch-and-Bound~\cite{montanaro_quantum_2020}.

Third, domain experts require abstraction software that allows them to turn their attention to the relevant parameters under their control and within their expertise: Modelling the problem, incorporating real-world constraints and objectives, and translating the solution of the mathematical model into the control language for their company processes. The amount of interdisciplinary expertise necessary to profit from quantum-enhanced solutions is significant and needs to be reduced through automation and abstraction.

The recent advances in QC hardware and improved accessibility have resulted in a wide variety of methods, tools and architectures aiming to provide abstractions to end users. High-level programming languages like Qrisp~\cite{seidel_qrisp_2024} or Qiskit~\cite{qiskit_contributors_qiskit_2023} simplify handling quantum circuits and even hybrid algorithms, translating them into lower-level representations like OpenQASM~\cite{cross_openqasm_2022} or QIR~\cite{QIRSpec2021}, and compiling and executing them. Specialized tools and solvers allow automating steps in the design and implementation process of a quantum-enhanced solution, e.g., by automating QC device selection~\cite{quetschlich_mqt_2025} or hyperparameter tuning for specific quantum algorithms~\cite{coelho_qaoa-predictor_2026}.

In existing tools, abstraction covers the mapping of a quantum algorithm to a QC hardware model~\cite{di_matteo_abstraction_2024}. With the QuaST Decision Tree (QDT), the intention is to extend the quantum software stack to fully reach the end user at the application level by automating (quantum and classical) algorithm selection and the tasks necessary for it -- problem encoding, hyperparameter tuning. Thus, the QDT goes beyond reducing implementation overhead for quantum algorithms. It provides a framework for \emph{decision support} which recommends combinations of quantum and classical algorithms suited to a specific use case. Automation tools with this purpose are currently constrained to specific types of quantum algorithms, e.~g., optimizing hyperparameters like the depth of a parameterized quantum circuit (PQC) in quantum approximate optimization (QAOA)~\cite{coelho_qaoa-predictor_2026}, or orchestrating a specific hybrid algorithm, e.~g. as Qiskit functions~\cite{noauthor_iskay_nodate, noauthor_optimization_nodate}. An overarching framework with the ability to provide guidance across a wide range of quantum-enhanced algorithmic options is not within their scope. To reach this kind of abstraction in a research field as active as quantum algorithm research, an \emph{extendable} framework is needed.

With this goal, we present the QuaST Decision Tree (QDT), a framework able to provide high-level abstractions for solving optimization problems, and demonstrate how automation can be reached with a module assessing the potential of concrete variational algorithms on specific problem instances. Our main contributions are
\begin{itemize}
    \item the design principles, requirements and architecture layout of the automation framework. An intermediate status of these principles is available in ~\cite{poggel_creating_2024}.
    \item the implementation of the automation framework in the QDT core.
    \item an automation module assessing the potential of a set of variational quantum algorithms (VQAs) by classifying the problem instance and using a scaling analysis~\cite{barligea_scalability_2025} from a database. 
\end{itemize}

The remainder of this paper is structured as follows: \Cref{sec:background} presents the necessary background and summarizes related work. \Cref{sec:design} details design principles and requirements for the flexible automation framework. Then, its realization and implementation in the QDT is described: its structure (\cref{sec:implementation}), main components (\cref{sec:components}), templates and input/output formats (\cref{sec:templates}) and usage scenarios (\cref{sec:usagescenarios}). \Cref{sec:scalabilitymodule} then describes the exemplary automation module based on a VQA scalability analysis~\cite{barligea_scalability_2025}. Finally, the implications and relevance of our contributions are discussed, including next steps in developing the application-facing abstraction layer (\cref{sec:discussion}).

\subsection{Notes on Implementation}

The implementations are made in Python (Version 3.11)~\cite{van_rossum_python_2009}. The notions of dictionaries, classes, instances, methods, attributes and objects are to be understood within the Python language (although a generalization would be possible). Abstract classes in this context refer to \emph{class templates} where the exact realization of some or all methods is left to so-called \emph{concrete} subclasses inheriting from the abstract class. 

\section{Background and Previous Work}
\label{sec:background}

For an overview on related work, \cref{sec:qcoptimization} starts with practical QC for optimization, and \cref{sec:abstractionsoftware} reviews existing approaches for application-facing abstraction. \Cref{sec:recapscalability} summarizes the foundation for the scalability-based automation module. Finally, \cref{sec:missingpieces} identifies the research gap addressed by the QDT and its modules.

\subsection{Practical Quantum Computing for Optimization}
\label{sec:qcoptimization}
Due to its enormous relevance for industry, combinatorial optimization (CO) is an intensively researched field for quantum-enhanced algorithms. In the QC context, the standard case is \emph{binary optimization}, in particular \emph{quadratic unconstrained binary optimization} (QUBO) formulation with $\vec{x^*} = \argmin_{\vec{x} \in \{0,1\}^n} \sum_{i,j=0}^{n-1} x_i Q_{ij} x_j$ where $Q$ is called \emph{QUBO matrix}. Many important CO problems can be mapped to QUBO, which can in turn be translated to an Ising model~\cite{lucas_ising_2014}.

However, finding a suitable formulation in terms of binary variables is nontrivial especially in the presence of complex constraints, e.~g. in vehicle routing~\cite{schnaus_efficient_2024} or scheduling ~\cite{schmid_highly_2025}.

The interest in applying QC to CO stems from its ability to explore large solution spaces in parallel, e.~g. in Grover's algorithm~\cite{grover_fast_1996} with a quadratic scaling advantage in computational complexity which is the best possible improvement for this problem type under standard assumptions~\cite{bennett_strengths_1997}. Whether such an advantage is enough to beat classical algorithms has to be investigated on a case-by-case basis for a specific combination of problem characteristics and algorithm~\cite{abbas_challenges_2024}. For this reason, applied research in QC-enhanced CO has led to a large variety of algorithms, typically using hybrid quantum-classical approaches for specific problem instances~\cite{abbas_challenges_2024}.

An important paradigm of hybrid QC for optimization are VQAs~\cite{cerezo_variational_2021}, especially variants of Quantum Approximate Optimization (QAOA)~\cite{farhi2014quantum, blekos_review_2024}. They use a quantum-classical iterative loop where a parameterized quantum circuit (PQC) repeatedly evaluates the loss function of the CO problem and a classical optimizer is tasked with finding the circuit parameters minimizing the loss function. However, they are plagued by trainability issues like barren plateaus~\cite{mcclean_barren_2018} and local minima~\cite{anschuetz_beyond_2022}. Specific issues have been addressed in countless proposed variants~\cite{cerezo_variational_2021}.

Beyond noisy QC devices, quantum-enhanced algorithms for CO with error-corrected QC have been proposed where practical benchmarking is expected to become possible in the next years, with approximately 200 logical qubits foreseen by several vendors~\cite{ibm_quantum_ibm_nodate, noauthor_ionq_nodate, noauthor_roadmap_2025}. Promising ideas include, e.~g., those based on Grover Adaptive Search, a variant of the Grover algorithm for CO~\cite{gilliam_grover_2021}, Quantum walk-based algorithms, and a quantum-enhanced Branch-and-Bound algorithm~\cite{montanaro_quantum_2020}. 

In terms of practical implementations, D-Wave's quantum annealer and hybrid solver offer solutions for hundreds and even thousands of binary variables~\cite{noauthor_advantage_nodate}. On gate-based systems, commercial software implements algorithms, typically for a specific collection of backends (e.~g., Qiskit functions for superconducting IBM backends~\cite{noauthor_functions_nodate}).

From an algorithmic side, central challenges are scaling to problem sizes with hundreds or thousands of variables which often requires the combination with classical decomposition methods, providing clear resource estimates for classical and QC resources, identifying promising applications, and managing the broad variety of algorithmic approaches outlined above to evaluate their practical benefits.

\subsection{Related Work in Abstraction Software}
\label{sec:abstractionsoftware}

Higher abstraction levels for QC are an active research topic~\cite{cobb_towards_2022, furntratt_towards_2023, di_matteo_abstraction_2024, di_matteo_art_2025}. Its primary focus is software development environments and high-level programming languages for convenient handling of quantum circuits. The diversity of applications, devices and stakeholders in the research field \cite{heim_quantum_2020} is reflected in a large variety of QC-specific programming languages, some linked to specific vendors and technologies like the widely used Qiskit for superconducting QC~\cite{qiskit_contributors_qiskit_2023}, others following a more independent approach, e.~g. Qrisp~\cite{seidel_qrisp_2024}. Common features are tools to define quantum circuits from individual gates or larger building blocks, with subsequent compilation for execution on hardware or simulated backends. Often, these steps, including the mapping to a hardware layout and circuit optimizations, are automated. However, end users still need the expertise to decide which algorithms and which quantum circuits to implement in order to fulfill their individual goals.

To support users in this regard, a recent push has come from design automation~\cite{zulehner_introducing_2020, soeken_programming_2018}. Tools are provided, e.~g., by the Munich Quantum Toolkit (MQT) comprising software for tasks like benchmarking, compilation, classical simulation, error correction, device selection and more~\cite{wille_mqt_2024}. The MQT Quantum Auto Optimizer~\cite{volpe_predictive_2024, volpe_towards_2024} is its user-facing abstraction tool for the CO domain. It employs supervised machine learning (ML) to select the best solver option from a given set based on QUBO input, combined with assistance in practical aspects like QUBO problem formulation. 

Another approach handles the complexity of the selection of a quantum-enhanced solution by systematic exploration of the options. This meta-solving approach maps the task of selecting a good solution strategy to an optimization problem itself~\cite{eichhorn_hybrid_2024}.

Apart from these and the earlier version of our QDT~\cite{poggel_recommending_2023}, no other approach exists, to our knowledge, for the challenge of selecting the best available algorithmic approach for CO problems from a general set (i.~e., not variants of a single algorithm). 

Beyond the circuit level, the commercial availability of quantum-enhanced solvers increases, as outlined in \cref{sec:qcoptimization}. Each relies on a specific algorithm for a fixed problem formulation (QUBO, HOBO, or Linear Programming). The hyperparameter tuning is then performed automatically. Thus, end users do not need to concern themselves with aspects such as the mapping problem of a quantum annealer~\cite{noauthor_advantage_nodate}, error suppression for a custom QAOA variant~\cite{sachdeva_integrated_2026}, or the discretization and counterdiabatic acceleration of an adiabatic solution protocol~\cite{chandaranaDigitizedcounterdiabaticQuantumApproximate2022}. While this greatly simplifies algorithm setup and hyperparameter tuning, it leaves the fundamental question of which algorithm to choose unanswered.

\subsection{Scalability Analysis of Variational Quantum Algorithms}
\label{sec:recapscalability}

A fundamental question for variational quantum algorithms (VQAs) is whether their resource requirements scale favorably enough to achieve practical quantum advantage, in particular taking into account the ability of the classical optimizer to find the optimum. A systematic method characterizes the measurement shot requirements of VQA/optimizer combinations~\cite{barligea_scalability_2025} and provides the empirical foundation for data-driven algorithm selection in the module presented in \cref{sec:scalabilitymodule}:

Two competing effects determine the feasibility of a VQA-optimizer combination for a given problem size~$n$: First, the finite sampling error $\varepsilon_{\mathrm{FS}}(n, n_{\mathrm{shots}})$ 
captures how measurement noise affects loss function evaluation, where $n_{\mathrm{shots}}$ denotes the number of measurement shots (circuit executions) used to estimate expectation values. This error scales as $1/\sqrt{n_{\mathrm{shots}}}$ for fixed system size, and grows exponentially in $n$ at fixed shot count. Second, the empirical optimizer tolerance threshold $\varepsilon^*(n)$ quantifies the maximum noise level under which the classical optimizer can still converge to a satisfactory solution. To extract it, optimizer success rates are measured and fitted across noise levels and converted to a normalized error measure using the variance of the CO loss function.

Crucially, the decline of $\varepsilon^*(n)$ with system size follows VQA- and optimizer-dependent scaling laws. Across the tested configurations, exponential, power-law, and logarithmic decay functions are reasonable hypotheses. Additionally, problem structure influences the fitted parameters, necessitating problem-aware analysis. By combining both effects, the solvability condition $\varepsilon_{\mathrm{FS}}(n, n_{\mathrm{shots}}) < \varepsilon^*(n)$ can be inverted to obtain the minimum shot requirement $n_{\mathrm{shots}}(n)$ for each VQA/optimizer combination under the different scaling hypotheses, with full error propagation providing uncertainty bounds on the estimates.

Finally, a hard feasibility criterion comes from the quantum disadvantage boundary: a quantum approach offers no benefit if its total computational cost $n_{\mathrm{shots}} \cdot n_{\mathrm{calls}}$ exceeds classical brute-force enumeration at $2^n$ evaluations. The number of quantum circuit calls $n_{\mathrm{calls}}$ depends on the optimizer. Any configuration where shot requirements exceed this boundary is outperformed by exhaustive classical search.

This methodology~\cite{barligea_scalability_2025}, thus, produces data for each analyzed VQA/optimizer combination: scaling parameters under all three hypotheses, shot requirement curves with propagated uncertainties, and feasibility classification across problem sizes. These results form the database underlying the automation module described in \cref{sec:scalabilitymodule}, enabling instance-specific resource estimation and algorithm recommendation within the QuaST Decision Tree.

\subsection{Missing Pieces for Reaching User-Level Abstraction}
\label{sec:missingpieces}

In conclusion, the following shortcomings require further advances, part of which are addressed in this work:
\begin{itemize}
    \item The high complexity of the selection, setup and tuning of a quantum-enhanced solution is an obstacle for ML-based solutions relying on large amounts of experimental data which is expensive to obtain. In this context, \emph{quantum-enhanced} refers to algorithms combining QC-based and classical subroutines, both individually solving meaningful problem parts.
    \item Applied QC is a highly active research field, producing many algorithmic variants. However, individual results often cannot be easily combined without systematic knowledge of their interplay.
    \item To realize the best solution for an optimization problem, the broad selection of algorithmic variants needs to be accessible through a common abstraction layer.
    \item The decision taking process needs to be made efficient, without the need for extensive benchmarking and trial-and-error approaches.
\end{itemize}

To address these issues, we build on the initial proposal of a recommendation framework for quantum-enhanced optimization~\cite{poggel_recommending_2023}. In comparison, we lay out a fundamental redesign of the QDT, replacing the layered architecture with a fully modular tree, including the Query mechanism to enable configurable automation. To emphasize the automation aspects, we demonstrate a concrete module to realize the vision of automated decisions based on a systematic scalability analysis of VQAs~\cite{barligea_scalability_2025}.
\section{Design Principles of the QuaST Decision Tree}
\label{sec:design}

From the QC hardware, abstraction is built up in layers, with more and more of the underlying complexity being shielded in order to allow users to focus on higher-level aspects. Thus, a quantum software stack is manifesting. When moving to higher abstraction levels and closer to the application, some challenges arise and become more pronounced:
\begin{enumerate}
    \item Non-technical personnel without expertise in the hidden complexity layers will increasingly come into contact with the quantum software stack. Even skilled developers (e.~g., at a company providing specialized software services) typically have no detailed knowledge on QC devices and algorithms, hence this issue is amplified. 
    \item With the broader target group of end users in various industry sectors, reaching consensus on unified or even standardized approaches and enforcing the use of specific workflows, templates, and formats becomes harder. The software increasingly needs to be able to deal with incomplete, faulty or imprecise input.
    \item  Requirements from outside the quantum software stack increasingly influence which interfaces it needs to provide. To generate a benefit, QC-enhanced optimization solutions will need to integrate into diverse business workflows and professional software environments, creating a wide set of industry-relevant usage scenarios with business-related requirements to the data flow, data availability, time constraints and desired output.
\end{enumerate}

At the inception of the QDT, the different processing steps from problem formulation to quantum algorithm execution were organized in layers: problem formulation and decomposition, encoding, algorithm selection, classical optimizer selection and compilation/hardware selection deferred to the lower-level software stack~\cite{poggel_recommending_2023}. Since this approach lacks flexibility and imposes impractical constraints on the available algorithms, the final QDT in this work features a fully modular setup for maximal flexibility and adaptivity. The backbone for this systematic approach remains a \emph{decision tree} structure, capturing and organizing the rich options to be made in the setup of a quantum-enhanced solution.

The vision of the QDT is to assist users by automating the decision between these options and providing recommendations optimized for application-specific quantum-enhanced algorithm performance. For sustainable benefits, the framework needs to be efficiently \emph{usable}, \emph{expandable} and \emph{maintainable}, covering the different usage scenarios in quantum-enhanced optimization (see \cref{sec:usagescenarios}). This vision leads to a set of design principles: 

\begin{enumerate}
    \item \emph{Automation without Restriction}: A convincing \emph{killer application}~\cite{hegde_beyond_2024} from the technology and business perspective is lacking. Therefore, abstraction software needs to stay open to innovation and avoid restricting the available exploration space for new algorithms and methods. Expert users should be able to understand the framework, retrace its decisions, and implement solution paths deviating from the recommendations.
    \item \emph{Modularity}: New modules can be inserted easily. The primary types of modules are automation subroutines, third-party software and adapters for the different usage scenarios.
    \item \emph{Locality}: To keep the maintenance of the framework feasible, adding or removing modules should require modifications only in modules that interact directly (e.g., by building on each other). Anticipating \cref{sec:implementation}, the locality will be realized by organizing the decision tree into \emph{nodes} performing the individual computing tasks. To capture the complexities and interdepencies in the decision tree, locality cannot be entirely fulfilled at every point -- for example, even high-level nodes often require information about the available hardware backends before deciding on a problem encoding.
    \item \emph{Configurability}: A central configuration source (e.g., a YAML file) should allow modifying the behavior of individual decision tree nodes and the full decision tree, notably controlling the degree of automation, and strategies used to derive recommendations.
    \item \emph{Efficiency}: The entire quantum-enhanced workflow realized by the framework needs to come with minimal overhead. In applications, the end-to-end performance metric will include the processing time within the framework.
    \item \emph{Transparency}: Data processing in the framework should be transparent, revealing what data is transmitted, where, how, and to what purpose it is modified.
\end{enumerate}

\section{Overview of the QuaST Decision Tree}
\label{sec:implementation}
The QDT is based on the design principles (\cref{sec:design}). It defines a framework that can be used for different purposes in the solution pipeline starting at the application. Its unique strengths are the ability to incorporate third-party software, and to adapt to its execution environment, be it a QC software stack at a HPC centre, a commercial solver, or standalone. The implementation in Python~\cite{van_rossum_python_2009} is a compromise to simplify scientific contributions at the expense of implementation efficiency. This shortcoming can easily be mitigated by leveraging, e.g., C-based modules for performance-critical components.

\begin{figure}[ht]
    \centerline{\includegraphics[width=0.95\columnwidth]{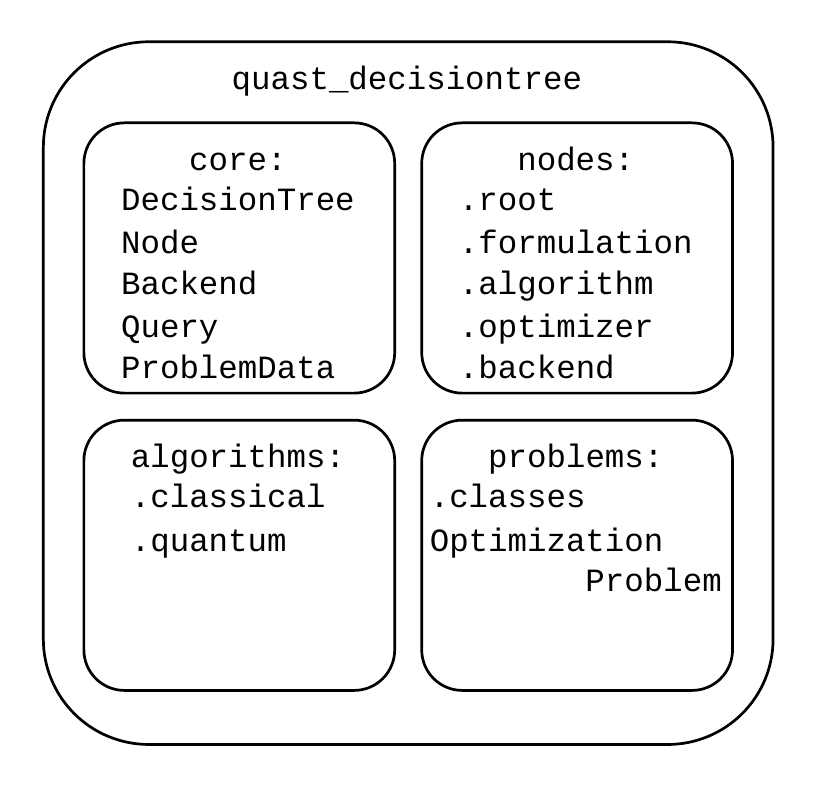}}
    \caption{Components of the \code{quast-decisiontree} package. \code{core} contains the framework, the \code{DecisionTree} base class and its building blocks. \code{nodes} provides a basic instance covering steps from problem loading to algorithm execution. \code{algorithms} and \code{problems} define templates for algorithms and problem classes, respectively.}
    \label{fig:structure}
\end{figure}

The QDT is provided in the \code{quast-decisiontree} package whose subpackage structure is shown in \cref{fig:structure}. The central Python object, the \code{DecisionTree} class and its components are contained in the \code{core} subpackage. The remaining subpackages \code{algorithms}, \code{nodes}, \code{problems} and \code{utils} provide necessary ingredients for a basic instance of the \code{DecisionTree} class which can illustrate and demonstrate its functionality while serving as a starting point for automation modules and concrete usage scenarios. The main automation and recommendation capabilities as well as adapters for different environments, e.g., the Munich Quantum Software Stack (MQSS)~\cite{burgholzer_munich_2026}, are then provided as plugins. One such automation module is described in \cref{sec:scalabilitymodule}. 

The \code{core} part comprises the following components, each realized as Python classes:

\begin{itemize}
    \item \code{DecisionTree} orchestrates the setup and execution of the QDT instances.
    \item \code{Nodes} are the constituents of the decision tree, able to flexibly execute atomic computational tasks.
    \item \code{ProblemData} is the centralized data structure passed between the nodes of the decision tree.
    \item \code{Queries} enable different levels of automation and serve as decision points. A global configuration flag enables either  the automated execution, or manual option selection.
    \item \code{Backends} are integrated via specialized nodes tasked with the communication with the backend (circuit definition, submission, and result retrieval).
\end{itemize}

The detailed functioning of the QDT follows in \cref{sec:components}, including the discussion on how the design principles from \cref{sec:design} are satisfied. A no-code instance is available online for exploration and testing~\cite{quastdecisiontree_website}.

From a high-level standpoint, the intended workflow with the QuaST Decision Tree is:

\begin{enumerate}
\label{enum:generalworkflow}
    \item Create an \emph{instance} of the QDT by specifying the nodes to include and their order.
    \item \emph{Configure} the behavior of the instance via a configuration file or by setting flags manually.
    \item \emph{Run} the QDT instance on the desired problem instance with the given configuration, by invoking its \code{run} method. This triggers a \emph{forward pass} through the decision tree, then reverses for a \emph{backward pass}. The typical breaking point between the passes is the submission and execution of the quantum algorithm, with algorithm setup and preprocessing on the forward pass and postprocessing, decoding and result interpretation on the backward pass.
\end{enumerate}

The output of the QDT depends on its configuration. One QDT instance can be used for multiple problem instances with the same configuration. E.g., an administrator can create and configure the QDT instance and allow users to execute runs. 

\section{Main Components of the QuaST Decision Tree}
\label{sec:components}

The detailed workflow realized by the QDT is laid out from the bottom up, first describing data handling and the vital components \code{Node} and \code{Query}, then arriving at the central orchestrator class \code{DecisionTree}.

\subsection{Data Handling}

The central data format of the QDT is a modified dictionary \code{ProblemData}, gathering all information for the current run. It is populated by the nodes, typically loading the problem instance first and subsequently adding additional processed data, e.g., a QUBO matrix or QASM-encoded quantum circuit. Depending on the key of a particular entry, low-level data validation is performed, e.g., a \emph{"qubo\_matrix"} entry should be a square two-dimensional array. The data is available in human-readable dictionary format, and nodes are required to indicate the entries they modify, which is essential in satisfying the transparency requirement.

\subsection{The Node class}

The atomic building block of the QDT and central enabler of modularity and locality is the \code{Node}. Nodes for specific tasks are inherited from the abstract \code{Node} class with attributes in \cref{tab:nodeclass}. When implementing a concrete \code{Node} for a specific task, the relevant \code{ProblemData} entries are defined, both the required ones and those that will be created or modified by the node. The QDT configuration then defines the children within the concrete QDT instance. During execution, the QDT instance makes all required \code{ProblemData} available. Furthermore, dynamic information can be requested during execution when necessary (e.g., queue lengths).

At runtime, \code{execute} and \code{interpret\_result} are invoked on the forward and backward passes respectively. For example, an encoding node translates an application-specific instance formulation to a QUBO format, and interprets the solution on the backward pass. 

The modularization of the QDT makes branching decisions challenging since the children of the nodes are only defined in a specific QDT instance, and nodes should function in different QDT instances. Specialized branching nodes therefore can define a set of known children, and select the correct one in their \code{next\_node} method. It is only invoked when a node has multiple children in a QDT instance. A typical example for branching are VQAs sharing parts of their setup such as the optimizer choice. A branching node can now realize non-local behavior. After the optimizer has been chosen, the algorithm setup and execution again depends on the particular VQA. A branching node will now read out the type of algorithm selected by an upstream node from \code{ProblemData}, and direct to the correct setup and execution nodes.

\begin{table}[ht]
\centering
\caption{Methods and Attributes of the Node class.}
\label{tab:nodeclass}
\begin{tabularx}{\columnwidth}{lp{2cm}p{4cm}}
\toprule
Method  & Important Parameters        & Description                                                                                                                                         \\ \midrule
\_\_init\_\_      & requires, creates, children & Executed on \code{DecisionTree} instantiation, defines necessary \code{ProblemData} entries and children. \\
execute           & problem\_data, path\_info   & Executed on the forward pass with the required problem data entries, and optional information to select a particular path.      \\
next\_node        &                             & Executed on the forward pass if the node has more than one child node.         \\
interpret\_result & result, problem\_data       & Executed on the backward pass with the result and the relevant problem data entries.                                              \\
request\_info     &                             & Pipeline to request data from the DecisionTree instance during runtime.            
\end{tabularx}
\end{table}

\subsection{The Query class}

Configurable automation is enabled by the \code{Query} class. A global flag sets the behavior of all queries in all nodes: On manual mode, users will be prompted for input, e.g., for algorithm selection. A default choice or recommendation based on the problem data can still be given. \code{Query} subclasses exist for various input types (various number formats, file paths, single-choice selection, multiple input values). Additional versatility is provided by combining queries into a \code{QueryTree} where subsequent queries can be executed conditionally on previous input. In the automatic mode, user prompts are skipped. Instead, default options, or recommendations computed from the problem data, are chosen. Queries are therefore essential to the requirement of automation without restriction.

\subsection{The DecisionTree class}

The overarching orchestration, data handling and execution is handled by the \code{DecisionTree} class whose important components are shown in \cref{tab:decisiontreeclass}. Its main functionalities are instance creation, configuration and execution.

\begin{table}[ht]
\centering
\caption{Methods and Attributes of the DecisionTree class.}
\label{tab:decisiontreeclass}
\begin{tabularx}{\columnwidth}{lp{2cm}p{4cm}}
\toprule
Method/Attribute & Important Parameters    & Description                                                                                         \\ \midrule
\_\_init\_\_     & config                  & Creates an instance based on the provided configuration file.                                       \\
validate         &                         & Validates the setup during initialization.                                                          \\
run              & problem\_instance, path & Execute a decision tree run (forward and backward pass) on the specified problem instance and path. \\
load\_backends   &                         & Retrieve all available backends from the specialized backend provider nodes.                  
\end{tabularx}
\end{table}

\begin{figure}[ht]
    \centerline{\includegraphics[width=1.05\columnwidth]{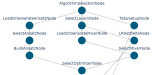}}
    \caption{The algorithm selection part of a basic \code{DecisionTree} instance. Based on the selected algorithms, different choices follow: the definition of an ansatz (e.g., variational quantum eigensolver~\cite{peruzzo2014variational}), a depth and mixer (QAOA~\cite{farhi2014quantum}), or skipping for a direct solver setup. Multiple nodes take part in the hyperparameter definitions. In the VQA case, the paths converge for the optimizer selection, which is not required in other cases (here: linear ramp QAOA~\cite{montanez-barrera_toward_2025} and a classical solver).}
    \label{fig:treeinstance}
\end{figure}

The instantiation of the \code{DecisionTree} class requires a configuration object in YAML file format or as a dictionary. It defines the source packages of concrete nodes (in the QDT core, this is \code{quast\_decisiontree.nodes}), the names of the nodes, their children, and optionally arguments for their initialization. By defining children, the entire tree structure is fixed. The required structure is a directed acyclic graph (DAG). Part of a QDT instance is shown in \cref{fig:treeinstance}.

After instantiation, a validation step matches \code{ProblemData} entries between nodes to verify that the setup is functional.

The \code{DecisionTree} instance maintains a dictionary of QC backends using a specialized \code{BackendNode}. This enables retrieving current backend information and submitting quantum circuits through an external submission pipeline. The backend nodes are adapters to enable abstraction across different access modalities. During execution, all nodes can then request information on available backends, with the \code{DecisionTree} instance requesting and forwarding the data.

Once the QDT instance is created, its structure (the nodes and their connections) is fixed, but the overall behavior can be configured -- e.g., how detailed the logging output is, and to what extent decisions should be taken without user input. Thus, the \code{DecisionTree} satisfies the configurability requirement.

The \code{run} method then executes the \code{DecisionTree}. A problem instance and further path specification can be passed directly. The problem instance in JSON format needs to match a corresponding problem class as described in \cref{sec:problemclass}, with a general QUBO class available next to application-specific instances. The path specification allows modifying the solution path taken through the decision tree, e.g., forcing an algorithm or setting a hyperparameter. To this end, node classes define \emph{path keys}: E.g., a QAOA node will accept a key setting the depth of the ansatz.

On execution, the QDT iterates through the nodes, starting from the root node and invoking the \code{execute} methods, making matching entries of the problem data, path specification, and on request dynamic information, available to the node. After node execution, the modifications are written back to the \code{ProblemData}, and execution continues with the next node. 

The forward pass finishes when a final node is reached, typically resulting in the execution of a hybrid algorithm. Once the result is available, the backward pass starts, iterating through the nodes in reverse for postprocessing, decoding and result interpretation. Similar to the \code{ProblemData} on the forward pass, a result dictionary is built on the backward pass. The backward pass then finishes with the root node which concludes the full run. In summary, the working of the QDT execution is shown in \cref{algo:decisiontree}.

\begin{algorithm}
\caption{QuaST Decision Tree}\label{algo:decisiontree}
\begin{algorithmic}[1]
\State config $\leftarrow$ load\_config()
\State node $\leftarrow$ init\_root\_node()
\State path $\leftarrow []$
\State problem\_data $\leftarrow$ init()
\While {node is not final}
    \State problem\_data $\leftarrow$ node.execute(problem\_data)
    \State append node to path
    \State node $\leftarrow$ node.next\_node()
\EndWhile
\State result $\leftarrow$ node.execute(problem\_data)
\For {node in reverse(path)}
    \State result $\leftarrow$ node.interpret\_result(result)
\EndFor
\end{algorithmic}
\end{algorithm}

\section{Templates, Input and Output Formats}
\label{sec:templates}

The \code{quast\_decisiontree} package contains templates for backends and problem classes, as well as a \code{Builder} mechanic to handle difference in the interfaces of, e.g., optimizers. The QDT uses YAML for configuration and tree definition, and JSON for problem instances and results. Run-specific and persistent logs are created automatically.

\subsection{Backends}

At the most basic level, QC backends accept a quantum circuit, execute it, and return a set of measurement outcomes. They are
typically embedded in a lower-level software stack handling compilation, optimization, and other pre- and postprocessing steps like error mitigation. The QDT aims to be device-agnostic and relies on specialized backend nodes to make the necessary data transformation to and from the backends. \code{DecisionTree} instances maintain a general list of available backends in a custom dictionary format, with mandatory specifications being a unique identifier and the provider node handling the backend communication. It is responsible for implementing the necessary data requests for specific information such as qubit counts, connectivity maps, calibration information etc.

\subsection{Problem Classes}
\label{sec:problemclass}

CO problem classes are defined by a specification of how to describe problem instances, and a way to evaluate the objective function. The QDT template \code{OptimizationProblem} additionally demands a \code{formulate\_problem} function used to translate the instance into a suitable format like QUBO. Problem instances are defined in JSON files, or equivalently Python dictionaries, with the application-specific definition in the \code{from\_dict} method. Different formulation options can be included via formulation modes. The objective evaluation is then required to be independent of the formulation, to encourage a clear, application-specific performance metric. For instance, a TSP class defines city-coordinate input, QUBO translation with selectable encodings, and a path-length objective. This interface allows the QDT nodes to load a problem instance and transform it for further processing. Additional problem classes can be integrated  without directly modifying the node loading the problem.

\subsection{Builders}

The \code{Builder} template allows the QDT to efficiently access options that are not available behind a common interface. The prototypical case is the selection of an optimizer in VQA setup. Optimizers provided, e.g., by scipy~\cite{virtanen_scipy_2020}, differ in the input to set them up: Gradient descent algorithms might require step lengths whereas genetic algorithms require a population size. 

For these cases, the \code{Builder} class provides a unified wrapper. \code{HyperParam} objects define the required input. Hence, a node can dynamically provide a range of options by collecting all builders of a certain type, obtaining a list of hyperparameters from the builders, and setting the required hyperparameters via user prompt, class defaults, or by invoking an automation module.

Builders are central for achieving modularity and expandability: After a node for a specific task such as optimizer selection is implemented, the range of options it provides can be expanded without modifying the node itself, simply by providing an appropriate builder instance. Currently, the core version uses builders for the creation of variational ansatzes according to a given template, and for optimizer selection. 

\section{Usage Scenarios and Validation}
\label{sec:usagescenarios}
The flexibility of the modular QDT is highlighted by the intended usage scenarios it can be adapted to, briefly described in this section. Initial validation is provided for two scenarios, with the commercial solver integration left as future work.

\subsection{Upper Part of an HPC Software Stack}

Within an HPC software stack handling quantum circuit and algorithm submission, the following setup is required:

\begin{itemize}
    \item A specialized backend provider node for circuit submission and retrieval of backend information.
    \item A well-defined QDT configuration from problem loading to the available algorithms and hyperparameter choices.
    \item A server infrastructure running the QDT instance, with submission formats mapping to the problem classes handled by the instance.
\end{itemize}

The QDT instance is then set up by an administrator, with free rein over the available algorithms, and what choices end users are allowed to make. From an end user perspective, only the preparation of the problem instance in JSON format and, optionally, the path file in YAML format is required. Then, the QDT instance translates the input into a format suitable for the lower HPC software stack, submits the circuits and processes the results. The end user profits from an \emph{extension of software stack capabilities}: Instead of submitting quantum circuits, they can directly submit their problem instances, trusting the automation of the QDT to provide the best solution.

As a first validation, a local instance of the QDT is used to submit circuits for Linear-Ramp QAOA~\cite{montanez-barrera_toward_2025} to the Munich Quantum Portal, offering access to QC backends via the Munich Quantum Software Stack~\cite{burgholzer_munich_2026}. Overhead and postprocessing times for random QUBO problems are shown in ~\cref{tab:validationmqp}. The QDT handles the encoding, preprocessing and decoding steps with minimal overhead when compared with the wait time for the results of circuit execution (which are dominated by the wait time between scheduling and execution of the jobs, even with an empty queue as in the experiment). Each instance was repeated 5 times.

\begin{table}[ht]
\centering
\caption{Execution overhead (wall-clock times) on the submission of LR-QAOA circuits to the 20-qubit backend \emph{EuroQExa20} via the Munich Quantum Portal}
\label{tab:validationmqp}
\begin{tabularx}{\columnwidth}{R{1.5cm}R{1.8cm}R{1.8cm}R{1.8cm}}
\toprule
QUBO size & Local QDT Forward Pass   & Wait Time for Result & Local QDT Backward Pass  \\ \midrule
4 & $< \qty{1}{\second}$ & $ \qty{11.8 \pm 1.7}{\second}$& $<\qty{1}{\second}$ \\
12 & $< \qty{2}{\second}$ & $ \qty{33.4 \pm 1.7}{\second}$& $<\qty{1}{\second}$ \\
20 & $< \qty{2}{\second}$ & $ \qty{39.4 \pm 1.5}{\second}$& $<\qty{1}{\second}$ \\
\bottomrule
\end{tabularx}
\end{table}

\subsection{Rapid Prototyping and Configuration}

In research and development, the QDT can be used to quickly configure experiments. The circuits are either run by the QDT instance itself or collected and submitted manually. The required setup is:

\begin{itemize}
    \item A well-defined QDT configuration containing the necessary solution paths (e.g., different algorithms, encoding choices or hyperparameters).
    \item Nodes implementing the novel paths to be explored.
    \item Path files defining the experiment configurations, e.g. different encoding choices, a set of benchmarking instances or hyperparameter selections.
\end{itemize}

The researcher maintains full control over the QDT configuration and the experiment workflow. The QDT simplifies the modification of the experimental setup and allows reproducing experiments. Furthermore, a specialized researcher does not need to implement the entire solution pipeline from scratch, but can use existing QDT nodes. Researchers benefit from being able to investigate the full solution pipeline in an applied setting, rather than isolated components.

For illustration, \cref{tab:prototyping_effort} shows the approximate lines of code (LOC) to integrate a novel solution component under the assumption that the component is available as a standard Python object (class or function). Once implemented, the end-to-end pipeline can be directly executed by providing a file specifying a QDT path. The implementation effort is particularly low when \code{Builder} classes are used, since no nodes need to be modified.

\begin{table}[ht]
\centering
\caption{Estimated integration effort for new components.}
\label{tab:prototyping_effort}
\begin{tabularx}{\columnwidth}{p{2cm}p{5cm}p{1cm}}
\toprule
Novel Component & Tasks for Integration   & Effort (LOC)                                                                                         \\ \midrule
Encoding for problem class     & add new formulation mode to problem class, and adapters reading in problem class information                  & \textasciitilde{}10                                       \\
Optimizer       & wrap optimizer in \code{Builder} class and define hyperparameters   &  \textasciitilde{}5             \\
Ansatz for variational algorithm & wrap ansatz generator in \code{Builder} class and define hyperparameters & \textasciitilde{}5 \\
Quantum algorithm & implement nodes for hyperparameter selection, and algorithm setup & \textasciitilde{}50 \\
Backend & implement node for backend circuit submission and information retrieval & \textasciitilde{}50 \\
Automation Module & read relevant information from problem data, insert into setup node, and characterize the selected option in problem data & \textasciitilde{}50  \\
Custom pipeline for information extraction & node including readout from result data, and executing the custom pipeline & \textasciitilde{}20 \\
\end{tabularx}
\end{table}

\subsection{Translation Component for Industrial Solvers}

Industrial solvers for optimization are deployed to end users in a customer-specific setup starting with data collection and preprocessing, then sending it to a commercial solver, and finally processing the result to provide the desired output. In this context, the QDT serves as a pre- and postprocessing accelerator for quantum-enhanced solutions. The manual overhead in deploying solutions for industrial optimization is reduced by including a QDT instance specialized in orchestrating hybrid quantum-enhanced solutions. Furthermore, this modularization makes already deployed solutions easily reusable and adaptable to new customers and solution pipelines. For providers of industrial solvers, this drastically reduces the effort in deploying quantum-enhanced solutions to end users.

\section{An Automation Module: Scalability of Variational Quantum Algorithms}
\label{sec:scalabilitymodule}

The purpose of the QDT is to provide abstraction by automation. To illustrate this, the automation module described in this section implements data-driven recommendations for VQA's. At its core sits a database of shot requirements $n_{\text{shots}}$ for various combinations of VQA and optimizer. Given a problem instance, the module then recommends the most resource-efficient combination, or provides an estimate for the user-selected configuration.

\subsection{Methodology and Database}
\label{sec:methodology}

The module is based on the rigorous methodology for determining the measurement shot requirements of VQA's~\cite{barligea_scalability_2025}. For a given VQA/optimizer combination, problem type, and QUBO density, it allows characterizing how the finite sampling error and the optimizer noise resilience threshold scale with system size, yielding a shot requirement curve $n_{\text{shots}}(n)$ and a comparison against the quantum disadvantage boundary (see \cref{sec:recapscalability}). The automation module uses a database with precomputed fit parameters, scaling exponents, and uncertainty estimates for the different configurations. The necessary data is obtained from precomputed benchmarking runs and can be continuously updated with new VQA/optimizer combinations and problem classes.

The database supports efficient queries, with the current coverage shown in \cref{tab:modulecoverage}. It can be continuously expanded and updated with new VQA variants, optimizers, problem classes, and combinations thereof. The extrapolated estimates to n = 100 naturally carry substantial uncertainty. The propagated error bounds quantify this. An expansion of the data base is planned to reduce the uncertainties.

\begin{table}[h]
\centering
\caption{Current coverage of the scalability database. For each configuration, $N=100$ random instances are sampled to ensure statistical robustness of the scalability analysis.}
\label{tab:modulecoverage}
\begin{tabularx}{\columnwidth}{lX}
\toprule
Category & Coverage \\
\midrule
VQA variants & VQE with hardware-efficient ansatz~\cite{peruzzo2014variational}, vanilla QAOA~\cite{farhi2014quantum}, warm-start QAOA~\cite{egger2021warm}, ma-QAOA~\cite{herrman2022multi}, QAOA+~\cite{chalupnikAugmentingQAOAAnsatz2022}, DC-QAOA~\cite{chandaranaDigitizedcounterdiabaticQuantumApproximate2022} \\
Classical optimizers & NGD~\cite{kuete2023universal}, COBYLA~\cite{powell1994direct}, SPSA~\cite{119632}, Powell~\cite{powell1964efficient}, NFT~\cite{PhysRevResearch.2.043158} \\
\midrule
\multicolumn{2}{l}{\textit{Problem-specific parameters:}} \\
\quad Random QUBO  & density $\rho \in \{0.1, 0.2, 0.3, 0.5, 0.7, 1.0\}$ \\
\quad MaxCut & density $\rho \in \{0.1, 0.2, 0.3, 0.5, 0.7, 1.0\}$ \\
\quad Knapsack & capacity ratio $r \in \{0.2, 0.4, 0.5, 0.6, 0.8\}$ \\
\midrule
System sizes (fitted) & $n = 3$ to $10$ \\
Extrapolation range & $n = 10$ to $100$ \\
\bottomrule
\end{tabularx}
\end{table}

\subsection{Functional Module Description}
\label{sec:io}

\textbf{Input.} The module requires a QUBO matrix $Q \in \mathbb{R}^{n \times n}$, supplied directly or via the QUBO problem class of the QDT. Then, its size $n$ (number of binary variables) and density (fraction of nonzero entries) are extracted.

Additionally, a problem class can be declared (e.g., Knapsack, MaxCut). In this case, the module uses the matching subset of the database to capture problem-specific scaling. Without a specified class, an automatic matching is performed based on size and density, using the general QUBO portion of the benchmarks. The user may then request either estimates for a specific VQA/optimizer combination or a recommendation among all available options.

\textbf{Output.} Module output is written to two files:
\begin{enumerate}
    \item \code{scalability\_assessment.json} contains a structured analysis with the input problem characteristics, shot estimates including uncertainty bounds under the three scaling scenarios, and the indication whether the estimate is \emph{feasible}, i.e., falls outside the quantum disadvantage regime. The VQA/optimizer combination with the lowest shot requirement is then recommended, if any fall below the quantum disadvantage threshold.
    
    \item \code{recommended\_config.yaml} provides the QDT configuration specifying the necessary node configurations including algorithm selection, optimizer setup, circuit depth, and backend selection. Hyperparameters are set to the values used in benchmarking and database creation, ensuring consistency between predicted shot requirements and execution.
\end{enumerate}

This dual output design lends itself to a fully automated operation mode where the configuration file is passed directly to the QDT, or manual inspection of analysis results.

\subsection{Architecture and Integration}
\label{sec:architecture}

The module contains four components:
\begin{itemize}
    \item The \textbf{QUBO Analyzer} extracts problem size and density from the input QUBO and handles problem type matching.
    \item The \textbf{Scaling Database} stores the scaling parameters obtained in previous benchmarking, organized by problem type, density, VQA variant, and optimizer.
    \item The \textbf{Shot Estimator} computes predictions of $n_{\text{shots}}$ under the three scaling scenarios.
    \item The \textbf{Recommendation Engine} evaluates the results, ranks combinations, and generates output files.
\end{itemize}

\textbf{Integration modes.} Depending on the usage scenario, the module is integrated into the QDT with one or two nodes:

\begin{enumerate}
    \item Recommendation mode: A node assessing scalability is inserted between problem encoding and algorithm selection. The QUBO matrix is read from \code{ProblemData}, then the VQA/optimizer combinations are evaluated. If feasible quantum combinations exist, the QDT proceeds with algorithm selection using the determined recommendations. If no quantum approach is feasible, the recommendation to choose a classical algorithm (or non-variational QC algorithm) is passed on.
    
    \item Estimation mode: A node to estimate scalability is inserted after algorithm selection. The node computes estimates for the chosen combination, allowing users to make informed decisions about resource allocation.
\end{enumerate}

\subsection{Demonstration on MaxCut} \label{sec:Demonstration}

As a demonstration, a MaxCut instance with $n=60$ variables and graph density $\rho=0.4$ is used. The mapping to QUBO is straightforward (see, e.g.,~\cite{barahona_experiments_1989}), therefore the module directly queries the database and computes shot estimates under three scaling hypotheses (exponential, power-law, logarithmic). To ensure robust recommendations, a conservative approach is adopted: for each configuration, 
the \emph{worst-case}, i.e., highest, estimate among valid fits is reported and compared to the quantum disadvantage threshold.

\Cref{tab:maxcut_results} shows the resulting assessment. The module classifies each configuration into one of three categories:
\begin{itemize}
    \item \textbf{Feasible:} The worst-case shot estimate falls below the 
          quantum disadvantage boundary.
    \item \textbf{Infeasible:} At least one scaling hypothesis yields a 
          valid estimate, but all exceed the boundary.
    \item \textbf{Not characterizable:} The VQA-optimizer combination 
          fails to produce consistent convergence behavior even at small 
          system sizes, preventing extraction of meaningful scaling 
          parameters under any hypothesis.
\end{itemize}

\begin{table}[h]
\centering
\caption{Worst-case shot estimates for a 60-variable weighted MaxCut 
instance (density $\rho = 0.4$). ``n.c.'' indicates non-characterizable configurations with inconsistent convergence.}
\label{tab:maxcut_results}
\renewcommand{\arraystretch}{1.4}
\begin{tabularx}{\columnwidth}{llXc}
\toprule
\textbf{VQA} & \textbf{Optimizer} & \textbf{$n_{\mathrm{shots}}$} & \textbf{Status} \\
\midrule
\multirow{4}{*}{VQE} 
    & NGD          & $1.9 \times 10^{6}$  & feasible \\
    & Powell       & $6.5 \times 10^{12}$ & feasible \\
    & NFT          & $1.9 \times 10^{15}$ & infeasible \\
    & COBYLA, SPSA & n.c. & -- \\
\midrule
\multirow{4}{*}{\shortstack[l]{ma-QAOA\\[-2pt]{\scriptsize $p=1$}}} 
    & Powell    & $5.8 \times 10^{14}$ & infeasible \\
    & COBYLA    & $1.4 \times 10^{22}$ & infeasible \\
    & NGD       & $8.3 \times 10^{51}$ & infeasible \\
    & NFT, SPSA & n.c. & -- \\
\midrule
\multirow{2}{*}{\shortstack[l]{QAOA+\\[-2pt]{\scriptsize $p=1$}}} 
    & Powell & $3.6 \times 10^{80}$ & infeasible \\
    & \shortstack[l]{NGD, NFT,\\[-2pt]SPSA, COBYLA} & n.c. & -- \\
\midrule
\multirow{3}{*}{\shortstack[l]{DC-QAOA\\[-2pt]{\scriptsize $p=5$}}} 
    & Powell         & $1.6 \times 10^{25}$  & infeasible \\
    & COBYLA         & $1.3 \times 10^{145}$ & infeasible \\
    & NGD, NFT, SPSA & n.c. & -- \\
\midrule
\multirow{4}{*}{\shortstack[l]{QAOA\\[-2pt]{\scriptsize $p=5$}}} 
    & Powell   & $9.1 \times 10^{19}$  & infeasible \\
    & SPSA     & $1.5 \times 10^{62}$  & infeasible \\
    & COBYLA   & $2.6 \times 10^{144}$ & infeasible \\
    & NGD, NFT & n.c. & -- \\
\midrule
\multirow{2}{*}{\shortstack[l]{WS-QAOA\\[-2pt]{\scriptsize $p=5$}}} 
    & NFT & $2.5 \times 10^{50}$ & infeasible \\
    & \shortstack[l]{NGD, SPSA,\\[-2pt]COBYLA, Powell} & n.c. & -- \\
\bottomrule
\end{tabularx}
\end{table}

Of all configurations evaluated, only VQE with the optimizers NGD and Powell produce feasible options. All QAOA-based variants, including vanilla QAOA, ma-QAOA, QAOA+, DC-QAOA, and WS-QAOA, either exhibit extreme shot requirements or cannot be characterized. Most QAOA variants fail to produce characterizable scaling behavior, 
reflecting the fundamental challenge posed by the rugged optimization landscapes that prevent reliable convergence even at small system sizes, making extrapolation to larger instances infeasible. The module's ability to identify and flag such cases is essential for trustworthy recommendations.
\balance
Based on this assessment, the module recommends VQE with NGD as the optimal configuration and generates the corresponding QDT configuration. If no feasible configuration exists for a given problem instance, the module routes to a classical solver, preventing resource usage on quantum-enhanced approaches that cannot outperform brute-force search.
\section{Discussion and Outlook}
\label{sec:discussion}

With the QDT, we present the concept and prototypical realization of a framework providing abstraction to end users of quantum-enhanced optimization. The central challenge is the realization of useful automation modules in a rapidly evolving, diverse environment. The QDT, therefore, is designed with maximum modularity and generality in mind to allow for \emph{automation without restriction}. The design principle outlined in \cref{sec:design} maximizes the usability of the QDT. \Cref{sec:implementation} and \cref{sec:components} then describe the framework following a tree-like structure of computation nodes acting on a transparent problem data format. Thus, seemingly contradictory goals are accomplished: hiding complex decisions behind an abstraction layer for end users while allowing experts to overwrite recommendations and explore novel paths. The resulting framework is useful in different situations (\cref{sec:usagescenarios}): in the context of an HPC software stack where it benefits end users by extending the software stack to their applications; as a tool for applied research and development, and as a translation layer to easily include quantum-enhanced solutions in commercial solvers.

A concrete automation module based on a robust VQA scalability analysis~\cite{barligea_scalability_2025}  shows how abstraction is provided by evaluating the problem-specific feasibility of VQA variants in conjunction with the classical optimizer.

With the QDT, quantum-enhanced solutions for CO are one step closer to adoption by end users. The design requirements lead to a framework useful to researchers and users alike. At the current maturity of the rapidly evolving field of applied QC research, this is crucial, for bridging the gap to industrial adoption. 

The crucial next development step in the QDT is its validation at increasing technology readiness, moving towards productive environments. The envisioned usage scenarios (\cref{sec:usagescenarios}) will lead to an iterative refinement and improvement of the framework. Additionally, further automation modules will be included in analogy to \cref{sec:scalabilitymodule}, or by integrating third-party software tools. 

\subsection*{Note On Code Availability}

The source code for the QuaST Decision Tree framework without the automation module described in \cref{sec:scalabilitymodule} will be provided as open source software in summer 2026.

\clearpage
\printbibliography

\end{document}